\documentclass[pre,aps,amsmath,amsfonts,amssymb,twocolumn,showpacs]{revtex4-1}
\usepackage{CJK}
\usepackage{graphicx}
\newcommand{\seq}{{\cal C}}
\newcommand{\addtag}{\refstepcounter{equation}\tag{\theequation}}
\begin{document}
\begin{CJK*}{UTF8}{mj}
\title{Phase transition in random adaptive walks on correlated fitness landscapes}
\author{Su-Chan Park (박수찬)}
\affiliation{The Catholic University of Korea, Bucheon 420-743, Korea}
\author{Ivan G.\ Szendro}
\affiliation{Institut f\"ur Theoretische Physik, Universit\"at zu K\"oln, K\"oln 50937, Germany}
\author{Johannes Neidhart}
\affiliation{Institut f\"ur Theoretische Physik, Universit\"at zu K\"oln, K\"oln 50937, Germany}
\author{Joachim Krug}
\affiliation{Institut f\"ur Theoretische Physik, Universit\"at zu K\"oln, K\"oln 50937, Germany}
\date{\today}
\begin{abstract}
We study biological evolution on a random fitness
landscape where correlations are introduced through a linear fitness gradient of
strength $c$. When selection is strong and mutations rare the
dynamics is a directed uphill walk 
that terminates at a local fitness maximum. We analytically
calculate the dependence of the walk length on the genome size
$L$. When the distribution of the random fitness component has an exponential tail
we find a phase transition of the walk length $D$ between a phase
at small $c$, where walks are short $(D \sim \ln L)$, and a phase at
large $c$, where walks are long $(D \sim L)$. For all other
distributions only a single phase exists for any $c > 0$.
The considered process is equivalent to a zero temperature
Metropolis dynamics for the random energy model in an
external magnetic field, thus also providing insight into the aging
dynamics of spin glasses. 
\end{abstract}
\pacs{87.23.Kg, 05.40.Fb, 75.10.Nr}
\maketitle
\end{CJK*}
\section{Introduction}
A population adapts to a new environment by accumulating beneficial
mutations. To study evolution in general and adaptation in particular, 
the picture of a fitness landscape has proven
to be helpful~\cite{deVisser2014}. 
Here a unique fitness value is assigned to each genotype which reflects the 
mean number of viable offspring
an individual with this genotype would produce. The mapping from genotype
to fitness defines the fitness
landscape. In this setting, adaptation is viewed as a
hill-climbing process that the population performs on the fitness
landscape. 

The interest in fitness landscapes has been renewed in
recent years as new techniques have made it possible to experimentally determine
the fitness for combinatorially complete sets of multiple genetic loci
\cite{deVisser2014,Szendro2012}. 
These experiments suggest that fitness landscapes typically contain a substantial amount of
randomness but also display correlations that smoothen the landscape. 
In particular, many empirical fitness landscapes possess multiple 
local maxima, i.e.\ genotypes with fitnesses that are higher
than those of all neighboring genotypes that can be reached by 
single-point mutations \cite{Whitlock1995}. Such local fitness peaks slow down adaptation 
by temporarily trapping the population, and it is important to
understand how long a population can evolve before it reaches a peak.

To address this question, we adopt the following simple but well
established model, which captures the evolutionary dynamics in a
regime of strong selection and weak mutation (SSWM)~\cite{G1983,G1984,O2002,O2005}.
Consider a population of $N$ individuals. Mutations occur with rate $\mu$, which is chosen small in the sense that
$N\mu\ll 1$. Selection is assumed to be strong enough that deleterious
mutations rapidly go extinct. If a beneficial mutation appears, 
it has a finite probability to become dominant in the population, and
this will happen before a second mutation can occur. Thus, in this regime, 
the whole population is almost always monomorphic, that is, genetically 
homogeneous. By means of sequentially fixing beneficial mutations, the population
``walks'' uphill through the fitness landscape, until it reaches a local fitness maximum, at which only deleterious mutations are available. Despite its
simplicity, the adaptive walk model has proven successful to describe
microbial evolution in the laboratory~\cite{RJCW2005,SCHOUSTRA2009,RAW2009}. 

A further common simplification is to suppose that all mutant genomes are of the same length $L$. Also, 
we only distinguish between genetic sites that are mutated and those that are not 
(diallelic loci, a common assumption in population genetics). This leads to an $L$-dimensional hypercubic genotype space of binary sequences
$\seq=(\dots,0,\dots,1,\dots)$, where zeros denote unmutated loci and ones
mutated loci. To assign fitness values to genotypes, we consider 
the rough Mount Fuji (RMF) model, which is a simple yet versatile model of tunably
rugged fitness landscapes~\cite{Aita2000,FKdVK2011,NSK2014} that has shown to
be capable of capturing many features of empirical fitness landscapes
\cite{Szendro2012,FKdVK2011,NSK2014,NSK2013}.
A realization of the landscape is constructed from independent
and identically distributed random variables $\eta_\seq$, which are
combined with a linear fitness gradient to define the genotype fitness as 
\begin{align}
\label{Eq:W}
W(\seq)=-c d(\seq,\seq_\mathrm{r}) + \eta_\seq.
\end{align}
Here the \textit{reference sequence} $\seq_\mathrm{r}=(1,1,\dots,1)$ has all loci mutated and
$d(\seq,\seq')$ is the Hamming distance between $\seq$ and
$\seq'$, with $d(\seq,\seq_\mathrm{r})$ being the number of zeros in $\seq$. 
The probability density of $\eta_\seq$ is $f(\eta_\seq)$ and the corresponding distribution function is $F(x) =
\int_{-\infty}^x f(y) dy.$ In the following we refer to $\eta_\seq$ as the \textit{random fitness
component} \cite{footnote1}.

When a walker is located at $\seq$, a further step $\seq\to\seq'$ is performed by 
choosing $\seq'$ at random with equal probability from the set
$\widetilde{\mathcal{N}}(\seq) = \{\seq' |W(\seq')>
W(\seq)\text{ and }  d(\seq',\seq) = 1\}$ of single mutant neighbors with higher fitness.
If this set is empty, $\seq$ is a local fitness maximum and the walker
stops. We refer to this dynamics as the random adaptive walk
(RAW) \cite{KL1987}. 
A key question in the theory of adaptive walks is the following~\cite{G1983,O2002,KL1987,Macken1989,FL1992,Jain2011,JS2011,NK2011,NSK2014}:
If the walker starts from the antipodal sequence
$\seq_a=(0,0,\dots,0)$ of $\seq_\mathrm{r}$, how many steps does it take before a fitness
maximum is reached and the walk terminates?
For the RAW on an uncorrelated random fitness landscape, corresponding
to the RMF model with $c=0$, the mean number of steps is known to be
$D_\mathrm{RAW} \approx \ln L + 0.099$ to leading order \cite{Macken1989,FL1992}. On the other hand, when
$c$ is much larger than the standard deviation of the random
fitness component in Eq.~\eqref{Eq:W}, the walker may take all $L$ steps
to the reference sequence with high probability. 

The purpose of this
paper is to clarify the nature of the transition between the regimes
$D_\mathrm{RAW} \sim \ln L$ and $D_\mathrm{RAW} \sim L$ that occurs as $c$ varies.
We show that a phase transition at an intermediate value of $c$ exists
if and only if the distribution of the random fitness component has an
exponential tail, and we characterize the transition in detail. 

The RAW arises from the full SSWM dynamics as an approximation when
fitness differences between neighboring genotypes are large \cite{Seetharaman2014}. 
The opposite case of small fitness differences has
been considered in \cite{G1984,O2002,Jain2011,JS2011,NK2011} for an uncorrelated landscape. 
We discuss the effect of using the full SSWM dynamics in
Sec.~\ref{Sec:fix}.

\section{\label{Sec:RAW} Random Adaptive Walks starting from the antipode}
\subsection{\label{Sec:form}Formal solution}
Our analysis starts from writing formally the probability density $q_l(\mathcal{Y}_l) \theta_l$ that an 
adaptive walker takes at least $l$ steps along a path $\mathcal Y_l$ before
it ends up at some local maximum. 
Here, $\mathcal{Y}_l$ is the ordered set of random fitness components $y_i$ of $\seq_i$ which have been visited by
the walker at the $i$'th step ($0\le i \le l$), 
$\mathcal Y_l\equiv (y_0,y_1,\ldots,y_l)$.  
We make the assumption that the distance to the reference sequence is strictly 
decreasing along the adaptive walk. 
Since the probability that a randomly chosen neighbor is located in the
direction of the reference state is $1- O(l/L)$, this assumption becomes exact as $L \rightarrow
\infty$ as long as the walk distance $l$ is $o(L)$. 
Within this assumption, the walker chooses
a random genotype from $\mathcal{N}(\seq_l) = \{\seq' \in \widetilde{\mathcal N}
|d(\seq',\seq_\mathrm{r})=d(\seq_l,\seq_\mathrm{r})-1\}$.
The condition that $y_{i-1}$ is smaller than 
$y_i + c$ for all $i=1,2,\ldots,l$ in $\mathcal Y_l$ will be called the walk condition and
$\theta_l$ is 1 (0) if the walk condition is (not) satisfied. 

Let us assume that the walker has taken $l$ steps to $\seq_l$ with the random fitness component 
$y_l$. 
Since the walker can choose any genotype from $\mathcal N(\seq_l)$,
the probability density of $y_{l+1}$ for a given $y_l$ is $f(y_{l+1})/(1-F(y_l-c))$
irrespective of the cardinality of $\mathcal N(\seq_l)$, as long as it is not zero.
Since $\mathcal N(\seq_l)$ is empty with probability 
$F(y_l-c)^{L-l}$, we get
\begin{equation}
q_{l+1}({\cal Y}_{l+1}) = 
f(y_{l+1}) \frac { 1 - F(y_l-c)^{L-l}}{1 - F(y_l-c)}q_l({\cal Y}_l) 
\theta_{l+1}.
\label{Eq:q}
\end{equation} 
We next define $Q_l(y_l,L)$ as the probability (density) to take $l$ steps and arrive at fitness $c(l-L)+y_l$. It is the integral of $q_l \theta_l$ over
all $y$'s but $y_l$,
$Q_l(y_l,L) = \int dy_0\cdots dy_{l-1} q_l({\cal Y}_l) \theta_l$,
and satisfies the recursion relation
\begin{align}
Q_{l+1}(y,L) = f(y) \int_{-\infty}^{y+c} Q_{l}(x,L)\frac{1- F(x-c)^{L-l}}{1 - F(x-c)} dx
\label{Eq:Qrecur}
\end{align}
with $Q_0(y,L) = f(y)$.
The probability $H_l$ that a walker takes at least
$l$ steps is obtained by integration over all endpoints
$H_l 
= \int_{-\infty}^\infty Q_l(y,L) dy,
$
and the probability $P_l$ that a walker takes exactly $l$ steps is
\begin{align}
\label{Eq:Pl}
P_l = H_l - H_{l+1} = \int_{-\infty}^\infty Q_l(y,L) F(y-c)^{L-l} dy.
\end{align}
Accordingly, the mean walk 
length can be calculated as
\begin{align}
D_\mathrm{RAW} = \sum_{l=1}^L l P_l=\sum_{l=1}^L H_l.
\end{align}

Although we have found a formal way of calculating $D_\text{RAW}$,
it seems very difficult to find an analytic solution for arbitrary $c$ and arbitrary $f(y)$ (see~\cite{FL1992} for the 
solution in the case of $c=0$).
Rather than directly analyzing Eq.~\eqref{Eq:Qrecur}, we use the following
approximation scheme. At first, we observe that
for $L\to\infty$ with $l$ kept finite, $H_l\to1$ (likewise $P_l\to 0$), and
$Q_l(y)\equiv Q_l(y,L=\infty)$ satisfies
\begin{equation}
Q_{l+1}(y)= f(y) \int_{-\infty}^{y+c} \frac{Q_{l}(x)}{1 - F(x-c)} dx,
\label{Eq:Ql}
\end{equation}
with $Q_0(y) = f(y)$.
According to Eq.~\eqref{Eq:Pl}, $P_l$ is almost 0 
as long as 
the region where $Q_l(y)$ is significant does not overlap with the region
where $F(y-c)^{L-l}$ is significant in the sense that the product $Q_l(y)F(y-c)^{L-l}\ll 1$ for all $y$. 
A way to determine whether the two regions overlap
is to check if $F(z_l-c)^{L-l}$ becomes of order unity, where $z_l$ is 
the mean of $Q_l(y)$, or 
\begin{align}
z_l \equiv \int_{-\infty}^\infty y Q_l(y)dy.
\end{align}
Once the two regions are significantly overlapped, they remain so for 
larger $l$ either
by decreasing $L-l$ or by increasing $z_l$, and $Q_l(y,L)$ becomes significantly smaller 
than $Q_l(y)$. Since $F(x)$ approaches 1 as $x$ gets larger and $F(z_l-c)^{L-l}$ can be significant
when $1-F(z_l-c) \sim O(1/(L-l))$, 
it suffices to estimate the solution of $F(z_l-c)^{L-l} = e^{-1}$ for an order of magnitude estimate of $D_\text{RAW}$.

\subsection{\label{Sec:exp}Exponential distribution}
We apply the above approximation scheme to the case of an exponential
distribution of random components, 
$f(x) = e^{-x}$, a common choice in the population genetics literature \cite{O2003}.  
After a substantial amount of algebra (see Appendix~\ref{Sec:appA}), we obtain
\begin{align}
Q_l(y) = 
 - \frac{d}{dy} \left ( \sum_{n=0}^l 
 y \frac{(y+cn)^{n-1}}{n!}  e^{-y - c n} \right ),
\label{Eq:Ql_exp}
\end{align}
and $z_l$ takes the form  $1 + \sum_{k=1}^l \xi_k$, with (see Appendix~\ref{Sec:app_sigma})
\begin{subequations}
\label{Eq:xil}
\begin{eqnarray}
\xi_l&=& z_l-z_{l-1}\\
\label{Eq:xi_int_large}
&=& \frac{(c l)^{l+1} e^{-cl}}{l!} \int_0^\infty  t e^{-ct}e^{(l-1)
g(t)}dt\\
&=& 1-c-\frac{(c l)^{l+1} e^{-cl}}{l!} \int_{-1}^0  t e^{-ct}e^{(l-1) g(t)}dt,
\label{Eq:xi_int_small}
\end{eqnarray}
\end{subequations}
where $g(t) =  \ln (1+t)-ct$.
Note that $g(t)$ has a unique (local) maximum at $t_M=(1-c)/c$, such that
it decreases (increases) for $t> t_M$ ($t<t_M$). 
In the case $c=1$, $\xi_l$ takes the simple form
\begin{equation}
\xi_l|_{c=1}
=  \frac{l^{l} e^{-l}}{l!}\sim \frac{1}{\sqrt{2\pi l}}.\label{Eq:xilcequal1}
\end{equation}

%%%%%%%%%%%%%%%%%%%%%%%%%%%%%%%%%%%%%%%%%%%%%%%%%%%%%%%%%%%%%%%%%%%%%%%%%%%%%%%%
%%%
\begin{figure}[t]
\includegraphics[width=\columnwidth]{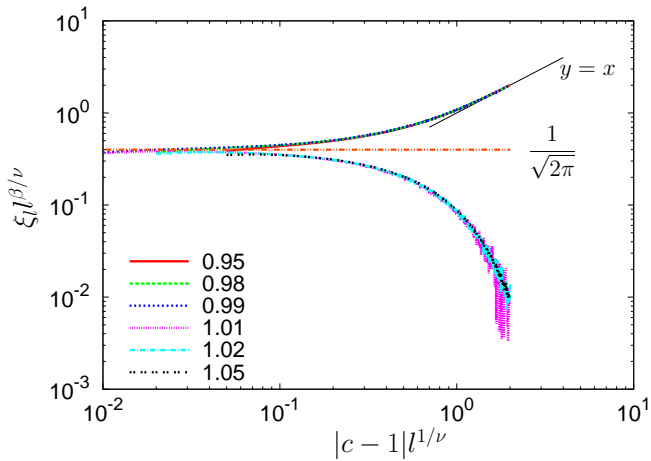}
\caption{
\label{Fig:collaps} (Color online) Scaling collapse plots of $\xi_l l^{\beta/\nu}$ vs $|c-1|
l^{1/\nu}$
with $\beta=1$ and $\nu=2$ for $c = 0.95, 0.98, 0.99, 1.01, 1.02$,
and 1.05. Horizontal line corresponds to $\psi(0)= 1/\sqrt{2\pi}$ and the 
slanted line shows $y=x$, which confirms the scaling ansatz
Eq.~\eqref{Eq:scaling}.} 
\end{figure}
%%%%%%%%%%%%%%%%%%%%%%%%%%%%%%%%%%%%%%%%%%%%%%%%%%%%%%%%%%%%%%%%%%%%%%%%%%%%%%%%
%%%%%%%%
For $c\neq 1$, we analyze the asymptotic behavior of $\xi_l$ for large $l$.
Since the integral domain in Eq.~\eqref{Eq:xi_int_large}
[\eqref{Eq:xi_int_small}] does not contain $t_M$ if $c>1$ [$c<1$], 
we use the Laplace method of asymptotic analysis, applying it to
Eq.~\eqref{Eq:xi_int_large} for the 
case of $c>1$ and Eq.~\eqref{Eq:xi_int_small} for $c<1$. When $l \gg 1$, the
main contribution
of the integral comes from the region around the maximum of $g(t)$ in
the integral domain. Since the maximum of $g(t)$  in the integral domain
of Eq.~\eqref{Eq:xi_int_large} [\eqref{Eq:xi_int_small}] for $c>1$ [$c<1$]
is at $t=0$, we approximate $g(t) \approx (1-c)t$,
which gives
\begin{align}
\xi_l \approx \text{max}(1-c,0) + \frac{e^{-l(c-1-\ln c)}}{\sqrt{2\pi l}} 
\frac{c}{(c-1)^2 l},
\label{Eq:xi_sol}
\end{align}
where we have used Stirling's formula.
Since $c-1-\ln c>0$ for $c\neq 1$, 
$\xi_l$ approaches $\text{max}(1-c,0)$ exponentially fast.
Also when $|c-1| \ll 1$, we can approximate
$c-1-\ln c \approx (c-1)^2/2$, 
suggesting a scaling form
\begin{equation}
\xi_l(c,l) = l^{-\beta/\nu} \psi((c-1) l^{1/\nu}),\label{Eq:scaling1}
\end{equation}
where, in the standard notation of critical phenomena, $\beta = 1$ and $\nu = 2$. 
Combining the approximations for the cases of $c\ne 0$ with
Eq.~\eqref{Eq:xilcequal1}, the asymptotic behavior of $\psi(x)$ takes the form
\begin{equation}
\psi(x) = 
\begin{cases}
1/\sqrt{2\pi}, & x \rightarrow 0,\\
e^{-x^2/2}/(\sqrt{2\pi} x^\nu), & x \rightarrow \infty,\\
|x|^\beta + e^{-x^2/2}/(\sqrt{2\pi} x^\nu), & x \rightarrow -\infty.
\end{cases}
\label{Eq:scaling}
\end{equation}
To confirm the scaling, we calculated $\xi_l$ for different values of $c$
using Monte Carlo simulations and the scaling plot is drawn
in Fig.~\ref{Fig:collaps}. We emphasize that the results 
of the Monte Carlo simulations are in complete agreement with
those obtained by
direct numerical integration of Eq.~\eqref{Eq:xil}.
Thus, we obtain
\begin{equation}
z_l = 1 + \sum_{m=1}^l \xi_m\sim
\begin{cases} (1-c) l, & c<1,\\
\sqrt{2l/\pi}, & c=1,\\
\text{finite}, & c>1.
\end{cases} 
\end{equation}

%%%%%%%%%%%%%%%%%%%%%%%%%%%%%%%%%%%%%%%%%%%%%%%%%%%%%%%%%%%%%%%%%%%%%%%%%%%%%%%%
%%%
\begin{figure}[t]
\includegraphics[width=\columnwidth]{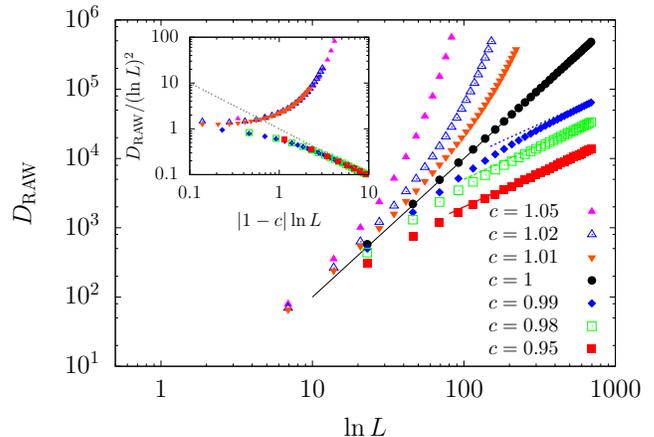}
\caption{
\label{Fig:Darw} (Color online) Plots of $D_\mathrm{RAW}$ vs\ $\ln{L}$ for $f(x)=e^{-x}$
and various choices of $c$ on a double-logarithmic scale. The black line shows $(\ln{L})^2$, while the other
lines correspond to $\ln{L/ (1-c)}$. (Inset) Scaling collapse plot of
$D_\text{RAW}/(\ln L)^2$ against $|1-c| \ln L$ on a double-logarithmic scale. The straight line with slope $-1$ is drawn to show the $\ln L$ behavior of
$D_\text{RAW}$ for $c<1$.
}  
\end{figure}
%%%%%%%%%%%%%%%%%%%%%%%%%%%%%%%%%%%%%%%%%%%%%%%%%%%%%%%%%%%%%%%%%%%%%%%%%%%%%%%%
%%%%%%%%
Since the distribution of the random fitness component is exponential, $Q_l(y)$ is
not expected to have a fat tail for large $l$. To confirm this expectation, we
calculated the standard deviation $\sigma_l$ of $Q_l(y)$ and found
that $\sigma_l \sim O(\sqrt{l})$ for $c \leq 1$ and $\sigma_l \sim
O(1)$ for $c > 1$; see Appendix~\ref{Sec:app_sigma}.
This implies that for $c<1$, $Q_l(y)$
can be well approximated by $\delta(y-z_l)$ for large $l$ and $P_l$ becomes
significant when
$l\sim (\ln L)/(1-c)$. For $c=1$, $z_l$ and $\sigma_l$ are
comparable
and $Q_l(y)$ cannot be approximated by a $\delta$ function. However, we expect that
when
$\ln F(z_l+\sigma_l) \sim -\ln L$, $P_l$ starts to become significant.
Hence, we conclude that
\begin{equation}
\label{Eq:darw}
D_\text{RAW} \propto \begin{cases} \ln L/ (1-c), & c<1,\\
(\ln L)^2, & c = 1,\\
O(L), & c>1.
\end{cases}
\end{equation}
In the limit $L \to \infty$ the ratio $D_\text{RAW}/L$ remains finite for $c>1$ but approaches 0 for $c\le
1$, which means there is a phase transition at the critical point $c^*=1$.
For $c=0$ we recover the result of \cite{FL1992}.
In Fig.~\ref{Fig:Darw} we compare our prediction to simulation results,
finding excellent agreement. Furthermore, Eq.~\eqref{Eq:darw} suggests that 
plots of $D_\text{RAW}/(\ln L)^2$ vs $(1-c) \ln L$ can be collapsed into a single 
curve, which is confirmed in the inset of Fig.~\ref{Fig:Darw}. Because the dynamics is invariant under the multiplication
of the fitness $W(\seq)$ by a constant factor, for a general
exponential distribution $f(x) = a^{-1} e^{-x/a}$ the critical point is given by the mean of
the distribution, $c^\ast = a$, and the walk length for $c < c^\ast$
is of the order of $\ln L/(1-c/a)$.

\subsection{\label{Sec:otherD}Other distributions}
Now we argue that the nature of the phase transition is determined
solely by the tail behavior of $f(y)$ and 
only exponential tails can induce a phase transition in
the large $L$ behavior as a function of $c$.
Let us revisit Eq.~\eqref{Eq:Ql}
and consider distributions $f(y)$ that are supported on the entire real axis.
Multiplying both sides of Eq.~\eqref{Eq:Ql} with $y$
and performing a partial integration, one can then derive the relation
%%%%%%%%%%%%%%%%%%%%%%%%%%%%%%%%%%%%%%%%%%%%%%%%%%%%%%%%%%%%%%%%%%%%%%%%%%%%%%%%
\begin{figure}
\includegraphics[width=\columnwidth]{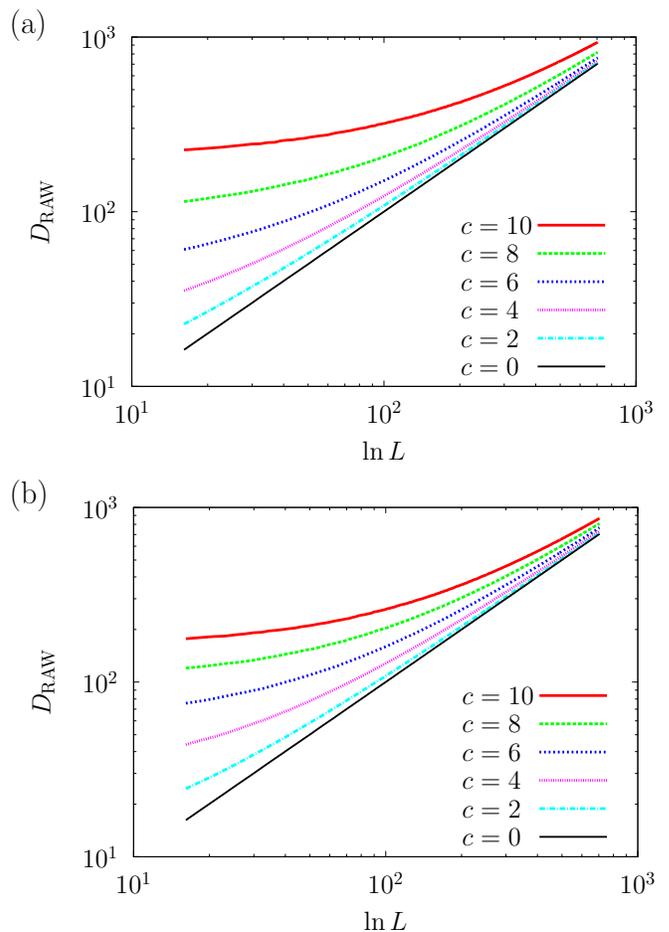}
\caption{
\label{Fig:DarwSE} 
(Color online) Double-logarithmic plots of $D_\mathrm{RAW}$ vs \ $\ln{ L }$ for various choices of 
$c$ 
(a) for the WD with $\alpha = \frac{1}{2}$ and (b) for the GPD with $\kappa = 0.5$.
The case of $c=0$, which is exactly solvable, is also drawn for comparison.
$D_\text{RAW}\sim \ln L$ in the large $L$ limit, independently of $c$.}
\end{figure}
%%%%%%%%%%%%%%%%%%%%%%%%%%%%%%%%%%%%%%%%%%%%%%%%%%%%%%%%%%%%%%%%%%%%%%%%%%%%%%%%
\begin{equation}
z_{l+1} - z_l = \int_{-\infty}^\infty \frac{Q_{l+1}(y)}{h(y)} dy - c,
\label{Eq:diffz}
\end{equation}
where $h(y)$ is the hazard function defined as
\begin{align}
h(y) \equiv \frac{f(y)}{1 - F(y)}.
\end{align}
Let us now \textit{assume} that $z_l\rightarrow\infty$ as $l\rightarrow \infty$
and that $Q_l(y)$ is reasonably concentrated, as was explicitly shown above
for the case when $f(y)$ is exponential.
Then we can replace the hazard function in the integral on the right hand side 
of Eq.~\eqref{Eq:diffz} with its asymptotic form for large arguments. 
Distributions with exponential tails are the only ones for which the hazard 
function approaches a constant for large $y$, specifically $\lim_{y\rightarrow
\infty} h(y) = a^{-1}$ for $-\ln f(x)=a^{-1} x + o(x)$. Inserting this into Eq.~\eqref{Eq:Ql} and using the fact that
$Q_l$ is normalized, we arrive at $z_{l+1}-z_l \approx a-c$, showing that 
$z_l \approx (a-c)l$ for $c < a$, while for $c > a$ the assumption that $z_l$ 
diverges is inconsistent. These results reproduce the previous
analysis for the purely exponential distribution [but note that in
this case the relation Eq.~\eqref{Eq:diffz} does not strictly hold, because the
support of the distribution is bounded on the left]. 

%%%%%%%%%%%%%%%%%%%%%%%%%%%%%%%%%%%%%%%%%%%%%%%%%%%%%%%%%%%%%%%%%%%%%%%%%%%%%%%%
\begin{figure}
\includegraphics[width=\columnwidth]{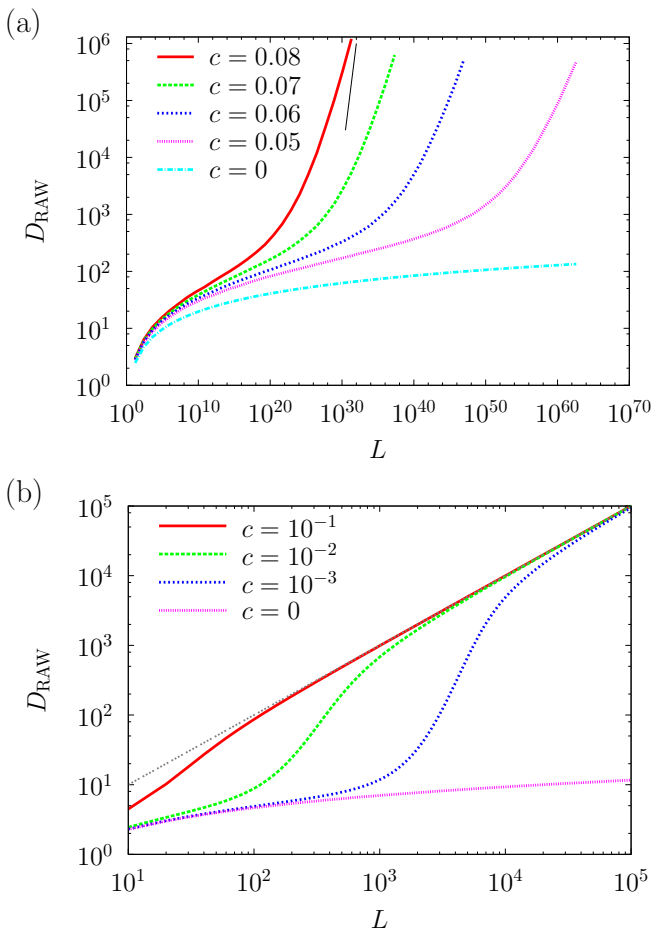}
\caption{\label{Fig:DarwGPD} 
(Color online) Double logarithmic plots of $D_\mathrm{RAW}$ vs \ $L$ for various choices of 
$c$ 
(a) for the WD with $\alpha = 2$ and (b) for the GPD with $\kappa = -1$.
The line segment in (a) and the straight line in (b) show the line $y\propto x$.
The case of $c=0$ which is exactly solvable is also drawn for comparison.
$D_\text{RAW}$ increases linearly in $L$ in the large $L$ limit, 
independently of $c$.}
\end{figure}
%%%%%%%%%%%%%%%%%%%%%%%%%%%%%%%%%%%%%%%%%%%%%%%%%%%%%%%%%%%%%%%%%%%%%%%%%%%%%%%%
For a tail of the form $\ln f(y)\sim -y^\alpha$, 
the asymptotic behavior of the hazard function is
$h(y) \sim y^{\alpha-1}$. Thus, the assumption that $z_l$ diverges is 
consistent only for $\alpha < 1$. Provided
$Q_l$ is sufficiently narrow we can estimate the integral on the right hand 
side to be of order $z_l^{1-\alpha}$; hence,
$z_{l+1} -z_l \approx z_l^{1-\alpha} -c$.
The asymptotic solution is $z_l \sim l^{1/\alpha}$ for any $c$, 
and it is straightforward to check that this implies that the walk length is always proportional to 
$\ln L$. Similarly, for a power law tail $f(y) \sim y^{-(\mu+1)}$ 
the hazard function $h(y) \sim \frac{1}{y}$, which leads to an 
exponential growth of $z_l$ for any $c$, and again to a walk length that is 
logarithmic in $L$. Conversely, for distributions with tails thinner than 
exponential such as the case $\alpha > 1$ mentioned above, the 
integral on the right hand side of Eq.~\eqref{Eq:diffz} never becomes large 
and the behavior is dominated by the negative term $c$ for any $c > 0$, 
leading to a walk length that is linear in $L$. 
Only when $c=0$ does one obtain $z_l \sim l^{1/\alpha}$, which implies that
the walk length is again $O(\ln L)$, consistent with the results in
\cite{FL1992}. Thus, we conclude that a
non-trivial transition is possible only for exponential tails.

To check our prediction that no phase transition occurs if the tail of the
distribution is not exponential, we numerically
calculated $D_\text{RAW}$ for the Weibull distribution (WD)
$F(x)=1-e^{-x^\alpha}$, with $\alpha=0.5$ and $\alpha=2$, and for the generalized Pareto distribution (GPD),
$F(x)=1-(1+\kappa x)^{-1/\kappa}$, with $\kappa=0.5$ and $\kappa=-1$. 
The case of $\kappa=-1$ corresponds to a uniform distribution.

As predicted, for the two cases where the tail of the distribution falls off
slower than exponentially, i.e.\ WD with $\alpha=0.5$ and GPD with $\kappa=0.5$,
$D_\text{RAW}$ will eventually, in the limit of large $L$, always grow as $\ln
L$, irrespective of the value of $c$; see Fig.~\ref{Fig:DarwSE}. For the distributions that fall off
faster than exponentially, i.e.\ WD with $\alpha=2$ and GPD with $\kappa=-1$, we
verify that $D_\text{RAW}$ grows as $\ln L$ for $c=0$ and as $L$
for any $c>0$; see Fig.~\ref{Fig:DarwGPD}. 
\section{\label{Sec:otherC}Generalizations}
In this section, we discuss three variants of the model. 
In Sec.~\ref{Sec:alpha}, we ask how changing the initial condition of
the RAW affects the phase transition point. 
To this end, we  abandon the assumption that the walker always
takes steps toward the reference genotype. 
In Sec.~\ref{Sec:non}, we  discuss how the phase transition
is modified when the linear fitness gradient in the RMF model is replaced
with a general nonlinear function of the distance to the reference sequence, focusing
on the case where the initial genotype is the antipode.
Finally in Sec.~\ref{Sec:fix}, we consider the full SSWM dynamics
where a step towards a fitter genotype, rather than occurring with
certainty, is accepted with a fixation probability $\pi_f$ that
depends on the fitness difference between the new and the old sequences.
Most of the discussion in this section parallels the arguments
in Sec.~\ref{Sec:otherD}.
For convenience, we use the same notation 
as in Sec.~\ref{Sec:form}
for similar quantities in this section.
\subsection{\label{Sec:alpha}Different initial condition}
Up to now, the initial genotype was taken to be the antipode of the reference
sequence. In this section, the walker is assumed to start from a 
genotype at Hamming distance ${\cal L}$ from the reference sequence,
where $0 \leq {\cal L} \leq L$.
When considering the infinite $L$ limit, the ratio ${\cal L}/L$
is kept finite; that is,
\begin{align}
\phi \equiv \lim_{L \rightarrow \infty} \frac{\cal L}{L}.
\end{align}
Note that the value of $\phi$ for the case considered in 
the previous section is $1$.

Suppose that the walker has already taken $l$ steps and the Hamming distance
of the $l$th genotype, say $\seq_l$, from the reference sequence is
$d(\seq_\mathrm{r}, \seq_l) = \ell$. Clearly,
there are $\ell$ neighbors in the direction towards the reference
sequence (the \textit{uphill} direction for short) and $L-\ell$
neighbors in the direction away from the reference sequence (the
\textit{downhill} direction) \cite{NSK2014}. 
Although at least one of the $L$ neighbors of the 
current genotype was encountered previously during the walk, the correlation
arising due to the previously assigned fitness value is negligible because
the probability that a mutation reverts to a previously observed
genotype is negligibly small as long as $L$ is very large \cite{FL1992,NSK2014}. Hence,
it is a good approximation to assume that the walker sees a new genotypic
environment after each step. Within this assumption, we can write 
a recursion relation similar to Eq.~\eqref{Eq:Qrecur}.

If the random part of $W(\seq_l)$ is $x$, 
the probability $P_\uparrow(n_1)$ that 
there are $n_1$ beneficial mutations in the uphill
direction and the
probability $P_\downarrow(n_{-1})$ that 
there are $n_{-1}$ beneficial mutations in the downhill direction are 
\begin{align}
P_\uparrow(n_1) &= 
\binom{\ell}{n_1}(1-F(x-c))^{n_1}
F(x-c)^{\ell-n_1},\\
P_\downarrow(n_{-1}) &= 
\binom{L-\ell}{n_{-1}} (1-F(x+c))^{n_{-1}} F(x+c)^{L-\ell-n_{-1}}.
\nonumber
\end{align}
Note that the probability of $n_1 = n_{-1} = 0$, which corresponds to 
the probability that the walker stops at $\seq_l$, is 
$F(x-c)^\ell F(x+c)^{L-\ell}$.
When there are $n_1$ and $n_{-1}$ beneficial mutations in the uphill
and downhill directions, respectively, the probability that
the walker takes a step toward the reference sequence [the antipode] is
$n_1/n$ [$n_{-1}/n$], where $n\equiv n_1+n_{-1}$. Hence the probability density
$\rho(y|x)$
that the random part of the next genotype is $y$ under the
condition that the walker will take a step is
\begin{align}
\rho(y|x) = \sum_{\sigma=\pm 1}\sum_{n=1}^L
\frac{n_\sigma}{n}
\frac{f(y)\theta(y-x+\sigma c)}{1-F(x-\sigma c)} 
P_{n_1,n_{-1}},
\end{align}
where the summation over $n$ stands for that over $n_1 = 0,\ldots, \ell$ and $n_{-1} = 0, \ldots,L-\ell$ with $n = n_1 + n_{-1} > 0$,
$\theta(x)$ is the Heaviside step function, and
$P_{n_1,n_{-1}} \equiv P_\uparrow(n_1)P_\downarrow(n_{-1})$.
Hence we get the recursion relation
\begin{align}
Q_{l+1}(y,L) = \int_{-\infty}^\infty \rho(y|x) Q_l(x,L)dx.
\end{align}

As in Sec.~\ref{Sec:form}, we now assume that $L$ is very large
and $l$ is small in the sense that $l/L \rightarrow 0$ and $\ell/L \rightarrow
\phi$ under the $L \rightarrow \infty$ limit.
Within this assumption, the probability distributions of
$n_1$ and $n_{-1}$ are sharply peaked around
$L\phi F(x-c)$ and $L(1-\phi) F(x+c)$, respectively.
Hence $Q_l(y) = \lim_{L \to \infty} Q_l(y,L)$ and its mean $z_l$
satisfy the recursion relations
\begin{align}
Q_{l+1}(y) =& f(y) 
\int_{-\infty}^\infty dxQ_l(x)
\nonumber \\
&\times \frac{\theta(y-x+c) + \varphi\theta(y-x-c)}{
 1 - F(x-c) +\varphi [ 1 - F(x+c) ]}
,\\
z_{l+1} - z_l =& \int_{-\infty}^\infty \frac{Q_{l+1}(y)}{h(y)} dy
\nonumber \\ & 
- c \int_{-\infty}^\infty dx Q_l(x)
\frac{1 - \varphi \tilde F(x,c)}{ 1 + \varphi \tilde F(x,c)},
\label{Eq:zalpha}
\end{align}
where $\varphi = (1-\phi)/\phi$ and $\tilde F(x,c) =
[1-F(x+c)]/[1-F(x-c)]$.
In the derivation of Eq.~\eqref{Eq:zalpha} it is implicitly assumed that 
the support of $f(x)$ extends over the whole real axis.
Note that when $\phi = 1$, the above equations reduce to Eqs.~\eqref{Eq:Ql} and
\eqref{Eq:diffz}, respectively. By symmetry, the case of $\phi = 0$
corresponds to a walker starting at the antipodal sequence with $c <
0$, and it is clear from the results of the previous section 
that the walk distance cannot be larger than for $c=0$.
So we restrict ourselves to the case of $c>0$ and $\phi >0$ in 
the following.

As in Sec.~\ref{Sec:otherD}, we first assume that 
$z_l$ diverges as $l \rightarrow \infty$ and that $Q_l(x)$
is highly peaked around $z_l$ for sufficiently large $l$.
When the tail is exponential, that is, $-\ln[1 - F(x)] = a^{-1}x + o(x)$
and $h(y) = 1/a + o(1)$, Eq.~\eqref{Eq:zalpha} for large $l$ becomes
\begin{align}
z_{l+1}-z_l \approx a - c \frac{1 - \varphi e^{-2c/a}}{1+ \varphi e^{-2c/a}}.
\end{align}
Hence the assumption that $z_l$ diverges breaks down if $c> a \tilde c$,
where $\tilde c$ is the (positive) solution of the equation
\begin{align}
\label{Eq:ctilde}
\varphi = \frac{\tilde c - 1}{\tilde c+1} e^{2 \tilde c}.
\end{align}
Thus, we conclude that the phase transition point depends on $\phi$.
When $\phi =1$, we get $\tilde c = 1$, as before, and when 
$\phi \ll 1$, $\tilde c$ diverges logarithmically with $\phi$ as 
$\tilde c \sim - \frac{1}{2} \ln \phi$.

\begin{figure}[t]
\includegraphics[width=\columnwidth]{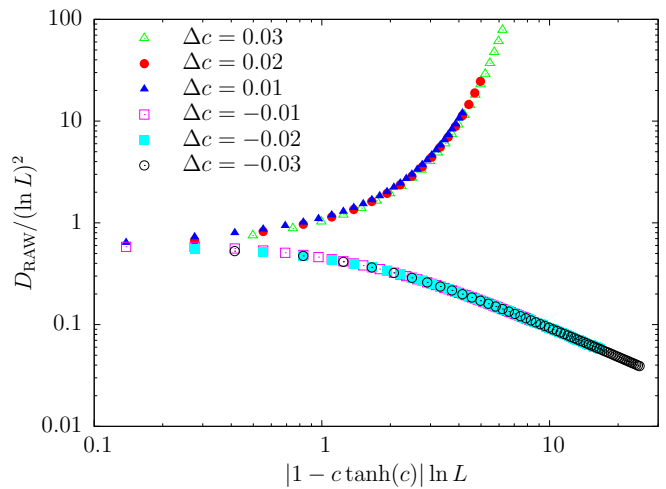}
\caption{\label{Fig:phi} (Color online) Finite size scaling collapse plot of $D_\text{RAW}/(\ln L)^2$ vs $| 1 - c \tanh (c) | \ln L$ for $c = \tilde c + \Delta c$ with
$\Delta c = 0.03$, 0.02, 0.01 (top data sets) and  $-0.01, -0.02, -0.03$
(bottom data sets), where $\tilde c =1.199$ 678 640 is the solution of
Eq.~\eqref{Eq:ctilde} with $\varphi = 1$.
}
\end{figure}
To support the above conclusion, we present simulation results for $\varphi = 1$ ($\phi = \frac{1}{2}$), with $f(x) = e^{-x}$ in Fig.~\ref{Fig:phi}.
The predicted transition point for $\varphi=1$ is determined
by the equation $1 - \tilde c \tanh(\tilde c) = 0$, whose solution is 
$\tilde c \approx 1.199~678~640$. Close to this value we expect a finite
size scaling collapse when $D_\text{RAW}/(\ln L)^2$ is plotted as a
function of $|1 - c \tanh (c) | \ln L$, which is indeed
the case as shown in Fig.~\ref{Fig:phi}.

The analysis for other distributions proceeds analogously to Sec.~\ref{Sec:otherD}.
If $-\ln[1 - F(x) ] = o(x)$ (slower than exponential decay), 
$\tilde F(z,c) \rightarrow 1$
as $z \rightarrow \infty$. Then Eq.~\eqref{Eq:zalpha} asymptotically
becomes Eq.~\eqref{Eq:diffz} with $c$ replaced with $c(2\phi - 1)$, 
which implies that the walk distance
is always $O(\ln L)$.
If $-1/\ln[1 - F(x) ] = o(1/x)$ (faster than exponential decay),
$\tilde F(z,c) \rightarrow 0$
as $z \rightarrow \infty$ and Eq.~\eqref{Eq:zalpha} asymptotically
becomes Eq.~\eqref{Eq:diffz}, which implies that the walk distance
is $O(L)$ as long as $c>0$.

To sum up, the initial condition of the RAW can affect the location of the critical point
for the case of distributions with an exponential tail, but does not
otherwise change the existence or nature of the phase transition.

\subsection{\label{Sec:non}Nonlinear deterministic fitness function}
The linear fitness gradient in Eq.~\eqref{Eq:W} implies that, in the
absence of the random fitness component $\eta_\seq$, each mutation
away from the reference sequence would decrease fitness by the same amount
$c$, and that the effects of different mutations combine additively.
However, in many cases it is observed that the effect of a mutation
depends on whether or not other mutations have occurred previously, a
phenomenon referred to as epistasis \cite{Chou2011,Khan2011}. 

To model such situations, we replace the linear deterministic part
in Eq.~\eqref{Eq:W} by a general function of the distance to the reference
sequence and ask how the phase transition is affected by
this modification. Since the main purpose 
of this section is to explain the qualitative change in the nature of
the transition, we restrict ourselves to the case
when the RAW starts at the antipodal sequence. As in Sec.~\ref{Sec:RAW}, we
assume that the walker always takes steps toward the reference sequence.

If the fitness of the sequence takes the form (recall that $\seq_a$ is the
antipode of the reference sequence)
\begin{align}
W(\seq) = k_{d(\seq,\seq_a)} + \eta_\seq,
\end{align}
it is straightforward to show that the recursion relation Eq.~\eqref{Eq:diffz} for
$z_l$ generalizes to 
\begin{align}
z_{l+1} - z_l =& \int_{-\infty}^\infty \frac{Q_{l+1}(y)}{h(y)} dy
 - \Delta k_l,
\label{Eq:zalpha_non}
\end{align}
with $\Delta k_l \equiv k_{l+1}-k_l$. In the following we assume that
$k_l$ is an increasing function of $l$ such that $\Delta k_l > 0$. As
explained in Sec.~\ref{Sec:form}, the walk length $D_\mathrm{RAW}$ will
be estimated from the solution of 
\begin{align}
F(z_l-\Delta k_l)^{L-l} = e^{-1}.
\label{Eq:Fest_gen}
\end{align}

To be concrete, let us consider distributions of the form $\ln f(y) \approx -a^{-1} y^{\alpha}$,
which gives $h(y) \approx a^{-1} \alpha y^{\alpha-1}$ for sufficiently
large $y$.
If we assume that $z_l$ diverges with $l$ and $Q_l(x)$ is well approximated 
by $\delta(x-z_l)$ for sufficiently large $l$,
Eq.~\eqref{Eq:zalpha_non} becomes
\begin{align}
z_{l+1}-z_l \approx \frac{1}{h(z_l)}- \Delta k_l \approx \frac{a}{\alpha} z_l^{1-\alpha} - \Delta k_l.
\label{Eq:dzdl}
\end{align}
Hence, the necessary condition for $z_l$ to diverge with $l$ is
$h(z_l) \Delta k_l < 1$, or $a z_l^{1-\alpha} >  \alpha \Delta k_l$.
If indeed $1/h(z_l) \gg \Delta k_l$ for sufficiently large $l$,
then the asymptotic form of Eq.~\eqref{Eq:dzdl} becomes 
\begin{align}
1 \approx h(z) \frac{dz}{dl} = - \frac{d}{dl} \ln (1-F[z(l)])
\end{align}
which gives 
\begin{align}
1 - F\left [z(l)\right ] \approx e^{-l}.
\label{Eq:hz}
\end{align}
Here $z(l)$ is an analytic continuation of $z_l$,
and it follows from the assumed shape of $F$ that 
$z(l) \sim l^{1/\alpha}$.
Since $\Delta k_l \ll 1/h(z_l) = O(z_l^{1-\alpha}) \ll z_l$ under the present
assumption, we can replace $F(z_l - \Delta
k_l)$ with  $F(z_l)$ in the condition Eq.~\eqref{Eq:Fest_gen} and it
follows from Eq.~\eqref{Eq:hz} that $D_\text{RAW} \sim O(\ln L)$.

To see when this scenario applies, we take $k_l$ to increase as
a power law \cite{Wiehe1997},
\begin{align} 
\Delta k_l = c l^{b-1} + o(l^{b-1}),
\label{Eq:Wiehe}
\end{align}
where $b=1$ corresponds to the linear fitness gradient. 
Then the condition $h(z_l) \Delta k_l \gg 1$ is 
fulfilled when $\alpha < 1/b$. That is, for distributions with the tail
decaying more slowly than $e^{-x^{1/b}}$, the mean walk distance is always
$O(\ln L)$ irrespective of the value of $c$. On the other hand,
if $\alpha > 1/b$, a trial solution $z_l \sim l^{1/\alpha}$ 
is contradictory to Eq.~\eqref{Eq:dzdl}, which suggests that the
walk distance is $O(L)$ for any $c>0$. 

\begin{figure}[t]
\includegraphics[width=\columnwidth]{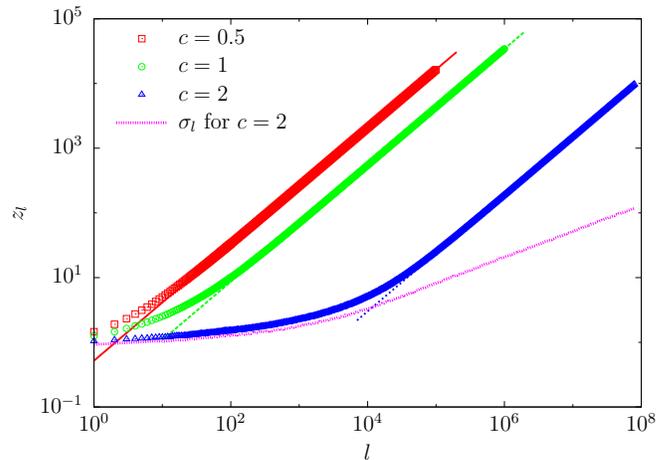}
\caption{\label{Fig:nonlin} (Color online) Double-logarithmic plots of 
mean $z_l$ (symbols) and standard deviation $\sigma_l$ (bottom line) 
of the distribution $Q_l(x)$ against $l$ 
for a nonlinear deterministic
  fitness function $\Delta k_l = c l^{b-1}$ ($k_0=0$) with $b = 0.9$. 
Lines show
  the asymptotic power law $z_l = (A l)^b$ with the prefactor $A$ given
  by the numerical solution of Eq.~\eqref{Eq:A}.
}
\end{figure}

In analogy with the linear case $b=1$, a possible 
phase transition is anticipated when $\alpha = 1/b$. In this case,
the asymptotic equation becomes
\begin{align}
z_{l+1}-z_l = a b z_{l}^{(b-1)/b} - c l^{b-1}.
\label{Eq:critical}
\end{align}
If we assume that $z_l \approx (A l)^\gamma$, the leading terms 
on both sides of Eq.~\eqref{Eq:critical} are consistent when $\gamma = b$,
and the prefactor $A$ satisfies the equation
\begin{align}
ab A^{b-1} - b A^b = c.
\label{Eq:A}
\end{align}
Inspection of Eq.~\eqref{Eq:A} reveals qualitatively different behaviors
for the cases $b > 1$ and $b < 1$, respectively. 
In fact, the case of $b>1$ turns out to require a different analysis
which is beyond the scope of this paper.
Hence we limit ourselves to $b<1$ and
defer the discussion about the case of $b>1$ to a future publication.

For $b < 1$, 
a unique positive solution of Eq.~\eqref{Eq:A} for $A$ can be found
for any $c > 0$, which implies that the walk length is always
logarithmic and a phase transition does not occur. To check the
validity of the assumptions leading to Eq.~\eqref{Eq:zalpha_non}, we have
determined $Q_l(x)$ and its moments by direct simulation. 
Figure~\ref{Fig:nonlin} strongly supports that 
$z_l$ asymptotically diverges as $l^b$ with the prefactor predicted
by Eq.~\eqref{Eq:A} 
for any $c > 0$ when $b < 1$.  Furthermore, it is clear from 
Fig.~\ref{Fig:nonlin} that the standard deviation $\sigma_l$ of $Q_l$ is negligibly 
small compared to $z_l$ for sufficiently large $l$, which 
supports the assumption that $Q_l(x)$ is well described by a $\delta$ 
function $\delta(x-z_l)$ in the asymptotic regime.

Quite generally, we see that the behavior of the walk distance is strongly
affected by the deterministic fitness profile and its interplay with the tail
of the distribution of the random fitness component. It is unclear at
present whether a phase transition as a function of $c$ is possible for
fitness profiles other than the linear one. 

\subsection{\label{Sec:fix}Finite fixation probability}
The probability of fixation of a beneficial mutation is a function
$\pi_f(s)$ of its selection coefficient, which in the present setting is simply the
fitness difference $s = W(\seq') - W(\seq)$ between the mutant
genotype $\seq'$ and the resident genotype $\seq$. The functional form
of $\pi_f(s)$ depends on the details of the underlying population
dynamics. For the particular case of Wright-Fisher dynamics, where
populations evolve in discrete generations and the number of offspring of an
individual is Poisson distributed \cite{Park2010}, the fixation probability is
well approximated by the expression $\pi_f(s) = 1 - e^{-2s}$
first derived by Kimura \cite{Kimura1962}. For small $s$ this reduces
to Haldanes classic result $\pi_f \approx 2s$ \cite{Haldane1927}, which is exact in
this limit, but for large $s$ the true fixation probability of the
Wright-Fisher model approaches unity somewhat more slowly, as $1 -
\pi_f(s) \sim e^{-s}$ \cite{Park2010}. For this reason we here use a slight
generalization of the Kimura formula, which reads 
\begin{align}
\pi_f(s) = 1 - e^{-\lambda s}.
\label{Eq:fixation}
\end{align}
For $\lambda \to \infty$ we thus recover the case of the RAW
studied in the previous sections, whereas for $\lambda \to 0$  
we obtain the Haldane-type fixation dynamics that is usually
considered in the SSWM literature \cite{G1983,G1984,O2002,O2005,Jain2011,JS2011,NK2011}. 

\begin{figure}[t]
\includegraphics[width=\columnwidth]{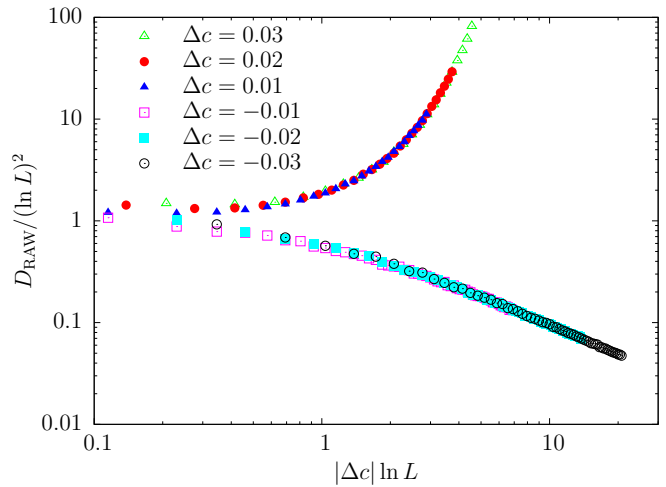}
\caption{\label{Fig:fixation} (Color online) Finite size scaling collapse plot of
  $D_\text{RAW}/(\ln L)^2$ vs $| \Delta c | \ln L$ for $\Delta c = c -
  \frac{4}{3} $, with
$\Delta c = 0.03$, 0.02, 0.01 (top data sets) and  $-0.01, -0.02, -0.03$
(bottom data sets). The fixation probability Eq.~\eqref{Eq:fixation} was
used with $\lambda = 2$, and the distribution of the random fitness
component is exponential with unit mean.
}
\end{figure}

As before, we consider the limit of infinite $L$.
In this case, we expect that effectively all possible values of the
random fitness components should appear with their appropriate weights.
Since the fixation probability of a beneficial mutation with
random component $y$ in the 
uphill direction is $\pi_f(y-x+c)$, where $x$ is the random 
component of the current genotype, we find the recursion relation 
for $Q_l(y)$ as \cite{Jain2011,JS2011,Seetharaman2014}
\begin{align}
Q_{l+1}(y) = \int_{-\infty}^{y+c} \frac{\pi_f(y-x+c) f(y)}{\int_{x-c}^\infty\pi_f(z-x+c) f(z) dz} Q_l(x) dx
\end{align}

Again, we are looking for a condition for $z_l$ to diverge. Assuming
that it diverges and that $Q_l(x)$ is highly peaked around $z_l$, we get
\begin{align}
z_{l+1} &= \int_{-\infty}^\infty dx \, Q_l(x) 
\int_{x-c}^\infty dy\frac{\pi_f(y-x+c) y f(y)}{\int_{x-c}^\infty\pi_f(z-x+c) f(z) dz} 
\nonumber \\
&\approx z_l - c + \int_{z_l-c}^\infty \frac{\tilde h(y)}{\tilde
  h(z_l-c)} dy
\label{Eq:zlfix}
\end{align}
where 
$\tilde h(y) \equiv \int_y^\infty \pi_f(z-z_l+c) f(z) dz$
and we have used $Q_l(x) \approx \delta(x-z_l)$.

One can readily evaluate the right hand side of Eq.~\eqref{Eq:zlfix} 
for the fixation probability Eq.~\eqref{Eq:fixation} and an exponential distribution
$f(x) = e^{-x/a}$, which gives
\begin{align}
z_{l+1} -z_l= a\frac{\lambda a +2}{\lambda a +1}-c.
\end{align}
This equation is consistent with a diverging solution for 
$c < a (\lambda a+2)/(\lambda a+1)$, and we conclude that the 
transition point is $c^\ast = a (\lambda a  +2)/(\lambda a +1)$. Note
that the $\lambda \rightarrow \infty$ limit reproduces the result
$c^\ast = a$ as anticipated, and $\lambda \rightarrow 0$, 
which corresponds to $\pi_f(x) \propto x$, gives $c^* = 2 a$.
Simulation results for $\lambda=2$ and $a=1$ are shown in
Fig.~\ref{Fig:fixation}. The simulations confirm that the transition
occurs at the predicted value $c^* = \frac{4}{3}$, and the nature of
the transition is the same as in the previously considered cases (compare to Figs.~\ref{Fig:Darw} and
\ref{Fig:phi}).

\section{Summary and discussion}
We have analyzed the mean adaptive walk length on random
fitness landscapes with a fitness gradient $c$ and various choices for
the distribution of the random fitness component. We showed that for distributions with
exponential tails, $D_\text{RAW}$
exhibits a continuous phase transition between a regime with
$D_\text{RAW}\sim\ln L$ for $c<c^*$ and $D_\text{RAW}\sim L$ for $c>c^*$.
For distributions that decay more slowly than exponentially, $D_\text{RAW}\sim L$ for
all $c>0$, and for distributions decaying faster than exponentially,
$D_\text{RAW}\sim \ln L$ for all choices of $c$. 

Note that the distinct role of the exponential distribution in
delimiting two regimes of qualitatively different behavior goes beyond the standard
classification in terms of extreme value theory \cite{deHaan2006}. Intriguingly, a
similar scenario appears in several other recent studies concerned
with records and extremes
\cite{Sabhapandit2007,Franke2012,Wergen2012}.
In the present context the special status of the exponential
distribution relies on the linear decrease of the deterministic
fitness profile with the Hamming distance from the
reference sequence (see Sec.~\ref{Sec:non}).

The mutational pathways followed by the RAW are
monotonically increasing in fitness, and a number of papers
have explored the conditions for the existence of such selectively
accessible paths \cite{FKdVK2011,Nowak2013,HM2014}. In particular, in \cite{HM2014} it
was proven that accessible paths to the reference sequence $\seq_r$ exist in 
the RMF
with a probability approaching unity for $L \to \infty$ and any $c >
0$. The present work shows, however, that the dynamic significance of such pathways
depends subtly on the tail properties of the fitness distribution, and
for heavy-tailed distributions they are essentially irrelevant for any
$c$. The tail also determines the behavior of the
number of maxima of the RMF landscapes for large $L$, which converge
to that of an uncorrelated random landscape for any $c>0$ when the tail
is heavier than exponential \cite{NSK2014}. 

Being a parameter of the fitness landscape, 
the strength of the fitness gradient $c$ governing the phase
transition cannot be easily tuned in an evolution experiment.
Nevertheless, the existence of two phases in which adaptive walk
lengths are proportional to $\ln L$ or $L$, respectively, is of 
considerable biological importance, because for realistic genome
sizes $L$ is vastly larger than $\ln L$. A recent numerical study addressing
the evolutionary benefit of recombination has found that these phases
persist also for genetically diverse populations where the SSWM
approximations do not apply \cite{NNSK2014}. As the advantage
of recombination is determined by how far a population can
adapt before being trapped at a local fitness maximum,
the existence of the phase of long adaptive walks shows that a
substantial advantage is possible even if the landscape is quite rugged.

Finally, we note that
the RAW dynamics considered in this paper is equivalent to a zero
temperature Metropolis dynamics \cite{FL1992}, where
genotypes $\seq$ are interpreted as configurations of $L$ spins 
with energies $-W(\seq)$ assigned according to the random energy
model in an external magnetic field $c$ \cite{D1981}. 
In that context we predict a novel kinetic phase transition as a function of field strength
from a low-field phase where the system gets stuck in a metastable
state after $O(\ln L)$ spin flips to a high-field phase where a finite
fraction of spins attain their ground state orientation. Our results
thus apply to aging processes in spin glasses, where rigorous analysis
has so far been restricted to the (less realistic) Glauber dynamics in
the absence of an external field and the energy distribution is always
assumed to be Gaussian \cite{bovier}.

\begin{acknowledgments} 
S.-C.P. acknowledges the support by the Basic Science Research Program through the
National Research Foundation of Korea~(NRF) funded by the Ministry of
Education, Science and Technology~(Grant No. 2011-0014680), 
by The Catholic university of Korea, Research Fund, 2014, and by the
University of Cologne within the Center of Excellence ``Quantum Matter
and Materials.'' J.K. acknowledges the kind hospitality of the Simons
Institute for the Theory of Computing, Berkeley,
during the completion of this work, and all authors acknowledge support
by Deutsche Forschungsgemeinschaft within SFB 680, SFB TR12, SPP 1590, and BCGS.
\end{acknowledgments} 
\appendix
\section{\label{Sec:appA}Derivation of Eq.~\eqref{Eq:Ql_exp}}
In this section, we show that Eq.~\eqref{Eq:Ql_exp}
solves the recursion relation Eq.~\eqref{Eq:Ql} for $f(y) = e^{-y}$.
Since
\begin{align}
&\int_{-\infty}^\infty Q_{l+1}(y) dy = \int_{-\infty}^{\infty} \frac{Q_l(x)}{1 - F(x-c)}
 dx \int_{x-c}^\infty f(y) dy \nonumber\\
&= \int_{-\infty}^{\infty} Q_l(x) dx = \int_{-\infty}^\infty Q_0(x) dx = 1,
\label{Eq:Q_normal}
\end{align}
for any $f(x)$, $Q_l(y)$ is normalized for any $l$ and for any $f(x)$.
Note that if $L$ is finite, 
\begin{align}
H_{l+1} &= \int_{-\infty}^\infty Q_{l+1}(y,L) dy \nonumber\\
&= \int_{-\infty}^{\infty} Q_l(x,L) \frac{1- F(x-c)^{L-l}}{1 - F(x-c)}
 dx \int_{x-c}^\infty f(y) dy\nonumber\\
&= H_l - \int_{-\infty}^\infty Q_l(x,L) F(x-c)^{L-l} dx,
\end{align}
which is Eq.~\eqref{Eq:Pl}

Since $1 - F(x-c) = e^{c-x}$ for $x>c$ and 1 for $x<c$,
Eq.~\eqref{Eq:Ql} can be rewritten as
\begin{equation}
Q_{l+1}(y) = e^{-y} \int_{0}^c Q_l(x) dx + e^{-y-c} \int_{c}^{y+c}  e^{x} Q_l(x) dx,
\label{Eq:recur_Q}
\end{equation}
with $Q_0(y) = e^{-y}$. One can easily find $Q_1$ and $Q_2$ such that
\begin{align}
e^y Q_1(y) =&   1 - e^{-c} + e^{-c} y ,\\
e^y Q_2(y) =&   \frac{1}{2} e^{-2 c} y^2+e^{-c} \left(e^{-c} (c-1)+1\right) y
\nonumber \\
&-c e^{-2 c}-e^{-c}+1 ,
\end{align}
which suggests that $Q_m(y)$ should take the form $e^{-y} j_m(y)$ with
\begin{equation}
j_m(y) = \sum_{k=0}^m \frac{a_{m,k}}{k!} y^k.
\end{equation}
This is a polynomial function of order $m$.
Since $Q_0(y) = e^{-y}$, $a_{0,0}=1$. 
Due to the normalization condition Eq.~\eqref{Eq:Q_normal}, the 
sum of $a_{m,k}$ over all $k$ for fixed $m$ should be 1; that is,
$\sum_{k=0}^m a_{m,k} = 1.$

From Eq.~\eqref{Eq:recur_Q}, we get
\begin{align}
j_{m+1}(y) 
= e^{-c} &\sum_{n=1}^{m+1}\frac{y^n}{n!} \sum_{k=n-1}^m a_{m,k} 
\frac{c^{k+1-n}}{(k+1-n)!}  \nonumber \\
&+ 
\sum_{k=0}^m a_{m,k}\left ( 1 - e^{-c} \sum_{n=0}^k \frac{c^{n}}{n!} \right ),
\end{align}
which yields a recursion relation for $a_{m,n}$ such that
\begin{align}
a_{m+1,0} 
&= 1 - e^{-c} \sum_{k=0}^m a_{m,k} \sum_{n=0}^k \frac{c^{n}}{n!},
\\ 
a_{m+1,n} 
&= e^{-c} \sum_{k=n-1}^{m} a_{m,k} \frac{c^{k+1-n}}{(k+1-n)! }.
\label{Eq:recur}
\end{align}
Note that 
\begin{align}
e^{c} &\sum_{n=1}^{m+1} a_{m+1,n}
= \sum_{n=1}^{m+1} \sum_{k=n-1}^m a_{m,k} \frac{c^{k+1-n}}{(k+1-n)!}\nonumber\\
&= \sum_{k=0}^m a_{m,k} \sum_{n=1}^{k+1} \frac{c^{k+1-n}}{(k+1-n)!}\nonumber\\
&= \sum_{k=0}^m a_{m,k} \sum_{s=0}^{k} \frac{c^{s}}{s!}
= e^c \left ( 1 - a_{m+1,0}\right ),
\label{Eq:norm1}
\end{align}
which again confirms the normalization condition Eq.~\eqref{Eq:Q_normal}.

To obtain $a_{m,k}$ for any $m,k$, we first find the explicit solutions
for $n=m$, $m-1$, $m-2$, and $m-3$ using Eq.~\eqref{Eq:recur} and then make 
an ansatz for $a_{m,k}$. Setting $n=m+1$, Eq.~\eqref{Eq:recur} becomes
$a_{m+1,m+1} = e^{-c} a_{m,m}$, which gives
$a_{m,m} =e^{-cm}$ with $a_{0,0}=1$.
Rewriting Eq.~\eqref{Eq:recur} as
\begin{equation}
e^{c(m+1)} a_{m+1,m+1-k} = e^{cm}a_{m,m-k} + e^{cm} \sum_{p=1}^{k} a_{m,m-k+p} \frac{c^{p}}{p!},
\end{equation}
which gives
\begin{equation}
e^{cl} a_{l,l-k} = e^{ck} a_{k,0} + \sum_{m=k}^{l-1} \sum_{p=1}^k e^{cm} a_{m,m-k+p} \frac{c^{p}}{p!},
\end{equation}
one can easily find $a_{l,l-k}$ after solving $a_{l,l-m}$ for
$m=0,1,\ldots,k-1$.
For example,
\begin{align}
a_{m,m-1} &= e^{-c(m-1)} + e^{-mc} ( (m-1) c - 1),\end{align}\begin{align}
a_{m,m-2} &= e^{-cm} \left ( c^2 \frac{m(m-2)}{2} - (m-1) c \right ) 
\nonumber \\ &+ e^{-c(m-1)} \left ( (m-2) c - 1 \right ) + e^{-c(m-2)},
\end{align}
\begin{align}
a_{m,m-3} &= 
\left ( \frac{m^2(m-3) }{6} c^3 
-\frac{m(m-2)}{2} c^2 \right ) e^{-cm} 
\nonumber \\
&+e^{-c(m-1)} \left(\frac{(m-1)(m-3)}{2} c^2 -c (m-2)\right)\nonumber 
\\&+e^{- c(m-2)} (c (m-3)-1)+e^{- c(m-3)}.
\end{align}
The above solutions of $a_{m,k}$ for specific $k$'s suggest the general form
\begin{align}
a_{m,0} &= 1 - \sum_{p=0}^{m-1} b_1^{m-p},
\label{Eq:amk1}
\\
\label{Eq:amk}
a_{m,k} &= \sum_{p=0}^{m-k} \left (b_k^{m-p} - b_{k+1}^{m-p} \right ),
\end{align}
where
\begin{align}
b_k^{n} = e^{-cn}\frac{n^{n-k-1} k}{(n-k)!} c^{n-k}
\end{align} 
with the convention $b_k^{n} = 0$ for $k> n$.
We first show that Eqs.~\eqref{Eq:amk1} and \eqref{Eq:amk} satisfy the normalization condition:
\begin{align*}
\sum_{k=1}^m a_{m,k} 
&
= \sum_{p=0}^{m-1} \sum_{k=1}^{m-p} 
\left (b_k^{m-p} - b_{k+1}^{m-p} \right )
=\sum_{p=0}^{m-1} b_1^{m-p},
\addtag \label{Eq:norm2}
\end{align*}
which combined with Eq.~\eqref{Eq:amk1} meets the normalization condition.
In the above calculation, we have changed the order of sum in such a way that
$\sum_{k=1}^m \sum_{p=0}^{m-k} = \sum_{p=0}^{m-1} \sum_{k=1}^{m-p}$.

Now we have to verify that Eq.~\eqref{Eq:amk} indeed solves
Eq.~\eqref{Eq:recur}.
To this end, it is convenient to use the identity
\begin{align}
A_{k,n}^{q}&\equiv b_k^{q} \frac{e^{-c} c^{k+1-n}}{(k+1-n)!}
= 
 \frac{q+1}{nq}b_n^{q+1}k B^{q+1-n}_{k+1-n}\left (\frac{1}{q+1}\right ),
\end{align}
where 
\begin{align}
B^N_n(x) = \binom{N}{n}(1-x)^{N-n} x^n.
\end{align}
Using
\begin{align}
\sum_{k=n-1}^q k B^{q+1-n}_{k+1-n}\left (\frac{1}{q+1}\right )
= \frac{nq}{q+1},
\end{align}
we get
\begin{align}
\sum_{k=n-1}^{m-p} A_{k,n}^{m-p} = b_n^{m+1-p}.
\end{align}
Finally, we can prove the validity of Eq.~\eqref{Eq:recur} as
\begin{align*}
&\sum_{k=n-1}^m \sum_{p=0}^{m-k} \left ( A_{k,n}^{m-p} - A_{k+1,n+1}^{m-p}
\right )\\
&=\sum_{p=0}^{m+1-n} \sum_{k=n-1}^{m-p} \left ( A_{k,n}^{m-p} - A_{k+1,n+1}^{m-p}
\right )\\
&=\sum_{p=0}^{m+1-n} \left ( b_n^{m+1-p} - b_{n+1}^{m+1-p} \right )
= a_{m+1,n},
\end{align*}
which is valid for $n\ge 1$. Since the case for $n=0$ is automatically
satisfied because of Eqs.~\eqref{Eq:norm1} and \eqref{Eq:norm2}, this completes the proof.

Using Eq.~\eqref{Eq:amk}, we can rewrite $j_m(y)$ as
\begin{align*}
j_m(y) &= 1 + \sum_{k=0}^m \sum_{p=0}^{m-k} \frac{y^k}{k!} \left (
b_k^{m-p} - b_{k+1}^{m-p} \right )\\
&= 1 + \sum_{n=1}^{m}\sum_{k=0}^{n} \frac{y^k}{k!} b_k^{n}
-\sum_{n=0}^{m}\sum_{k=0}^{n} \frac{y^k}{k!} b_{k+1}^{n},
\addtag
\end{align*}
where we have changed the order of sums and we have set $n = m-p$.
Since
\begin{align}
\sum_{k=0}^{n} \frac{y^k}{k!} b_k^{n}
&= e^{-cn} \frac{y(cn)^{n-1}}{n!n} \sum_{k=0}^n k \binom{n}{k}  \left (
\frac{y}{cn}\right )^{k-1} \nonumber\\
&= \frac{(cn+y)^{n-1}}{e^{cn} n!} y,
\end{align}
and 
\begin{align}
\sum_{k=0}^{n} \frac{y^k}{k!} b_{k+1}^{n}
&= \frac{e^{-cn} c (cn)^{n-2}}{(n-1)!} \sum_{k=0}^{n-1}
(k+1)\binom{n-1}{k} \left ( \frac{y}{nc} \right )^k \nonumber\\
&= e^{-cn } 
\frac{(y+cn)^{n-2}}{(n-1)!} (y+c),
\end{align}
we get
\begin{align}
j_m(y) = \sum_{n=0}^m 
\frac{(y+cn)^{n-2}}{n! e^{cn}} \left [
y^2 + (c-1) ny - nc \right ],
\end{align}
which gives
\begin{align}
Q_l(y)
&=\sum_{n=0}^l
\frac{(y+cn)^{n-2}}{n! e^{cn+y}} \left [
y^2 + (c-1) ny - nc \right ] \nonumber\\
&=
- \frac{d}{dy} \left ( \sum_{n=0}^l 
 y \frac{(y+cn)^{n-1}}{n!}  e^{-y - c n} \right ),
\label{Eq:Qsol}
\end{align}
which is Eq.~\eqref{Eq:Ql_exp}.
By substitution, one can easily check that Eq.~\eqref{Eq:Qsol} indeed solves
Eq.~\eqref{Eq:recur_Q}.

\section{\label{Sec:app_sigma}Mean and standard deviation of $Q_l(y)$}
In this section, we calculate the mean $z_l$ and the standard deviation
$\sigma_l$ 
of $Q_l(y)$. For convenience we introduce $\xi_l$ and $\Xi_l$, which are defined
as
\begin{align}
\label{Eq:xi}
\xi_l &= \int_0^\infty y \left [ Q_l(y) - Q_{l-1}(y) \right ] dy,\nonumber\\
\Xi_l &= \int_0^\infty y^2 \left [ Q_l(y) - Q_{l-1}(y) \right ] dy.
\end{align}
Obviously,
\begin{equation}
z_l = 1 + \sum_{m=1}^l \xi_l, 
\quad \sigma_l = \left ( 2 + \sum_{m=1}^l \Xi_l -z_l^2\right )^{1/2} .
\end{equation}
After an integration by parts, we obtain
\begin{align*}
\xi_l &= \int_0^\infty  y \frac{(y+cl)^{l-1}}{l!}  e^{-y - c l}dy
\\
%&= \frac{(c l)^{l+1} e^{-cl}}{l!} \int_0^\infty  t (1+t)^{l-1}  e^{-c lt}dt \\
&= \frac{(c l)^{l+1} e^{-cl}}{l!} \int_0^\infty  t e^{-ct}e^{(l-1) ( \ln (1+t)-ct)}dt
\end{align*}
which is Eq.~\eqref{Eq:xi_int_large} and
\begin{align*}
\xi_l&= 1 - c - \int_{-cl}^0 y \frac{(y+cl)^{l-1}}{l!}  e^{-y - c l} dy \\
%&=1-c- \frac{(c l)^{l+1} e^{-cl}}{l!} \int_{-1}^0  t (1+t)^{l-1}  e^{-c lt}dt \\
&= 1-c-\frac{(c l)^{l+1} e^{-cl}}{l!} \int_{-1}^0  t e^{-ct}e^{(l-1) ( \ln
(1+t)-ct)}dt,
\end{align*}
which is Eq.~\eqref{Eq:xi_int_small}. Likewise, we get
\begin{align*}
\Xi_l &= 2 \int_0^\infty  y^2 \frac{(y+cl)^{l-1}}{l!}  e^{-y - c l}dy \\
%&= 2 \frac{(c l)^{l+2} e^{-cl}}{l!} \int_0^\infty  t^2 (1+t)^{l-1}  e^{-c lt}dt \\
&= 2 \frac{(c l)^{l+2} e^{-cl}}{l!} \int_0^\infty  t^2 e^{-ct}e^{(l-1) ( \ln (1+t)-ct)}dt
\addtag
\end{align*}
which is suitable to analyze for $c>1$.
For $c<1$, it is convenient to analyze
\begin{align}
\Xi_l=& 2l(1-c)^2+2-2 \frac{(c l)^{l+2} e^{-cl}}{l!} 
\nonumber
\\
&\times \int_{-1}^0  t^2 e^{-ct}e^{(l-1) ( \ln (1+t)-ct)}dt.
\end{align}

Since
\begin{align}
y \frac{(y+l)^{l-1}}{l!}  e^{-y -  l}  &= - \frac{d}{dy} \left [ \frac{(y+l)^l}{l!} e^{-y-l}
\right ],\nonumber\\
y^2 \frac{(y+l)^{l-1}}{l!}  e^{-y -  l}  &= - \frac{d}{dy} \left [ y \frac{(y+l)^l}{l!} e^{-y-l}
\right ] \nonumber\\
&+ \frac{(y+l)^l}{l!} e^{-y-l},
\end{align}
$\xi_l$ and $\Xi_l$, for $c=1$, become
\begin{align}
\xi_l|_{c=1} &= -\int_0^\infty \frac{d}{dy} \left [ \frac{(y+l)^{l}e^{-(y+l)}}{l!}\right ] dy 
= \frac{l^l e^{-l}}{l!},\\
\nonumber\\
\Xi_l|_{c=1} &= 2 \int_0^\infty \frac{(y+l)^l}{l!} e^{-y-l} dy\nonumber\\
&= 2\frac{l^{l+1}e^{-l}}{l!} \int_0^\infty e^{l(\ln(1+t)-t)} dt\nonumber\\
&\sim 2 \sqrt{\frac{l}{2\pi}} \int_0^\infty e^{-lt^2/2} dt = 1,
\end{align}
where we have used $\ln(1+t) - t \approx -t^2/2$ for small $t$.

Using the same method to arrive at Eq.~\eqref{Eq:xi_sol}
for the asymptotic behavior of $\xi_l$, we get for $c<1$
\begin{align}
\Xi_l \sim
\displaystyle 2 l (1-c)^2 + 2 - 2 \frac{l^l e^{-l}}{l!} e^{-l(c-1-\ln c)} \frac{c^2}{l(1-c)^3}
\end{align}
and for $c>1$
\begin{align}
\Xi_l \sim
 2 \frac{l^l e^{-l}}{l!} e^{-l(c-1-\ln c)} \frac{c^2}{l(1-c)^3}.
\end{align}
To sum up, we obtain
\begin{equation}
z_l \sim \begin{cases}
(1-c) l, & c<1,\\
\sqrt{2 l/\pi}, & c=1,\\
\text{finite}, & c>1.
\end{cases}
\quad
\sigma_l \sim \begin{cases}
O(\sqrt{l}), & c\leq1,\\
\text{finite}, & c>1.\\
\end{cases}
\end{equation}

\section{Numerical measurement of $D_\text{RAW}$}
In order to verify our analytical predictions and check whether they are
still valid when we lift the restriction that RAWs should only move
towards the reference sequence, we
performed numerical simulations. These were carried out as follows. Before the
first step, the population is positioned at ``height'' $h=0$ and it is assigned a
fitness value $W_0=\eta_0$, where $\eta_0$ is drawn from the considered
distribution $f(\eta)$. By height we mean the Hamming distance from the antipodal sequence. For each step, a new neighborhood consisting of $L$
states is drawn. To each of the $L-h$ states in the forward direction
(at height $h+1$), a fitness value is assigned, which is drawn according to $W_i=(h+1)c+\eta_i$. Correspondingly, the
$h$ backwards neighbors (at height $h-1$) obtain fitness values according to
$W_i=(h-1)c+\eta_i$. To speed up the simulations, the fitnesses are assigned to
the neighboring states in a random order until a fitness value larger than the
one selected after the last step is generated. Then the population is
transferred to the corresponding state and the height is updated.
The walk terminates when there are
no neighbors satisfying the condition on the fitness. The walk length $D_\text{RAW}$ is
estimated by averaging the number of steps performed up to this point,
$n_\text{steps}$, over ensembles of RAW's. For the data presented here, we
considered ensembles of $10^3$ to $10^5$ walks. Note that, in order to be able to
simulate large landscapes, previously encountered fitness values and the information about which
states are neighbors are not stored. However, for large $L$ this should not
considerably alter the measured values of $D_\text{RAW}$.

%%%%%%%%%%%%%%%%%%%%%%%%%%%%%%%%%%%%%%%%%%%%%%%%%%%%%%%%%%%%%%%%%%%%%%%%%%%%%%%%
\begin{figure}[t]
\includegraphics[width=\columnwidth]{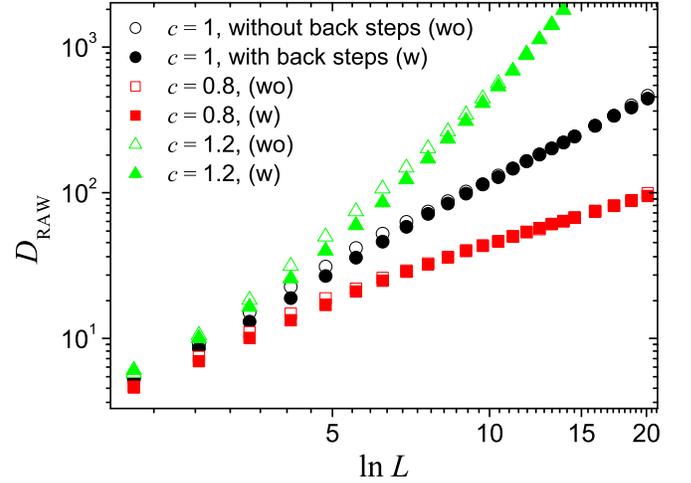}
\caption{
\label{Fig:Darwwobs} (Color online) Comparison of $D_\text{RAW}$ vs \ $L$ for simulations
with and without back steps. As can be verified, curves show
excellent agreement for $\ln L > 10$ ($L > 2 \times 10^5$).} 
\end{figure}
%%%%%%%%%%%%%%%%%%%%%%%%%%%%%%%%%%%%%%%%%%%%%%%%%%%%%%%%%%%%%%%%%%%%%%%%%%%%%%%%
In Fig.~\ref{Fig:Darwwobs}, we compared simulations with (w) and without (wo) backward steps 
for the case of $F(x) = 1 - e^{-x}$ for various choices of $c$. All curves show
excellent agreement for sufficiently large $L > 10^6$, which shows that our
analytical results remain valid for the original model that includes back
steps. 

If $L$ is extremely large (note that the largest $L$ in Fig.~\ref{Fig:Darw} is $10^{300}$), even deciding a fitness of the first step by
the above procedure is infeasible because we have to generate $L$ random
numbers. Therefore, direct simulation of Eq.~\eqref{Eq:q} is used to
simulate RAWs without back steps.
The algorithm is as follows: Assume that the walker is located
at ``height'' $h$.
Since the walker can take a next step with probability $P_\text{walk} 
= 1-F(x-c)^{L-h}$,
a single random number generation is necessary to decide whether it stops there.
When calculating $F(x-c)$ for very large $x$, one should be very careful
if $1-F(x-c)$ is smaller than the machine accuracy. For example, if one uses
a double precision calculation, $1-F(x-c)$ should be larger than $10^{-16}$;
otherwise, $1 - [1-F(x-c)]$ will be regarded as $1$ by a computer, which
gives $1 - F(x-c)^{L-l} = 0$. (Note that $( 1 - 10^{-20})^{10^{200}}$ is almost
zero but careless computation will give 1.) In case $1-F(x-c)$ is very small [in our simulations,
``very small'' means  $1-F(x-c) < 10^{-10}$], we approximate $P_\text{walk}$ as
\begin{align}
P_\text{walk} &= 1 - e^{(L-h) \ln \left (1-[1-F(x-c)]\right )} \nonumber\\
&\approx 1 - e^{-(L-h) [1-F(x-c)]}.
\label{Eq:Pwalk}
\end{align}
Once the next step is determined to be taken, we generate a random number $y$
from the distribution ($y>x-c$),
\begin{equation}
\frac{F(y) - F(x-c)}{1 - F(x-c)}.
\end{equation}
In practice, we generate a uniformly distributed random number $z$, then we determine
$y$ by 
\begin{equation}
1 - F(y) = (1-z) \left [ 1 - F(x-c) \right ].
\label{Eq:Fyz}
\end{equation}
For $F(x) = 1 - e^{-x^\alpha}$,
\begin{align}
y = \begin{cases}
\left [ (x-c)^\alpha - \ln (1-z) \right ]^{1/\alpha}, & x>c,\\
\left [ - \ln (1-z) \right ]^{1/\alpha}, & 0<x<c,
\end{cases}
\end{align}
and for $F(x) = 1 - (1+\kappa x)^{-1/\kappa}$ ($\kappa > 0$),
\begin{align}
y=\begin{cases}
\displaystyle \frac{1}{\kappa} \left ( \frac{1+\kappa(x-c)}{(1-z)^\kappa} - 1 \right ), & x>c, \\
\displaystyle \frac{1}{\kappa} \left ( \frac{1}{(1-z)^\kappa} - 1 \right ), & x<c.
\end{cases}
\end{align}

To obtain $z_l$ via simulations, all we have to do is to get $y$ 
from Eq.~\eqref{Eq:Fyz} without considering $P_\text{walk}$ 
in Eq.~\eqref{Eq:Pwalk}.

\bibliographystyle{apsrev}
\bibliography{Park}

\begin{thebibliography}{33}
\bibitem[{foo({\natexlab{a}})}]{footnote1}
\bibinfo{journal}{Note that the random fitness components remain unchanged
  during the adaptive walk; that is, the random variables are quenched.}
\end{thebibliography}
\end{document}